\documentclass[titlepage,twoside,12pt]{article}
\usepackage{amssymb}
\usepackage{amsfonts}
\begin{document}

{\LARGE \bf Further on Pilot-Wave Theories} \\ \\

{\bf Elem\'{e}r E ~Rosinger} \\ \\
{\small \it Department of Mathematics \\ and Applied Mathematics} \\
{\small \it University of Pretoria} \\
{\small \it Pretoria} \\
{\small \it 0002 South Africa} \\
{\small \it eerosinger@hotmail.com} \\ \\

\hspace*{6cm} {\it Dedicated to Marie-Louise Nykamp} \\ \\

{\bf Abstract} \\

In [2], a detailed argument is presented on a version of pilot-waves, given by at Theory of Exclusively Local Beables. What the author of [2] considers
to be his crucial proposal is described in the title of section 3 as 'complicated, ugly, and highly contrived'. The reason for such a proposal is
claimed to be the widespread perception among quantum physicists, according to which 'those sympathetic to the pilot-wave ontology no doubt expected
that, for a system of N particles moving in three spatial dimensions, the theoretical description would be of N wave-particle pairs - each pair
consisting of a point particle guided in some way by an associated wave propagating in 3-space. But Schroedinger's wave function for such an N-particle
system was emphatically not a set of N (interacting) waves, each propagating in 3-space. It was, rather, a single wave propagating in the 3N-dimensional
configuration space for the system', [2, p. 5]. In the present paper it is argued that the mentioned widespread expectation need not be seen as being
betrayed by what actually happens. In particular, the construction in section 3 of [2] may be avoidable. \\ \\ \\

{\bf 1. Concepts and Beables} \\

{\bf 1.1. Concepts} \\

It is a fundamental feature of Physics, as well as of other sciences, the inevitability of the use of certain undefined
basic concepts. \\
Such a state of affairs is often overlooked due, among others, to tacit reference to well accepted intuitive meanings one
associates with such undefined concepts. Yet that fact remains, it has been known for longer, and one of those who warned
us about it was Newton himself in the Principia, where he keeps mentioning that he does not know what gravitation is,
except how it behaves, that is, proportionally with two given masses, and inverse proportionally with the square of the
distance between them. \\
More generally, Leibniz formulates the situation in his 'Of an Organum or Ars Magna', as follows : 'Whatever is thought by
us is either  conceived through itself, or involves the concept of another. So one must either proceed to infinity, or all
thoughts are resolved into those which are conceived through themselves. Every idea is analyzed perfectly only when it is
demonstrated a priori that it is possible. Since however it is not in our power to demonstrate the possibility of things
in a perfectly a priori way, that is, to analyze them into God and nothing, it will be sufficient for us to reduce their
immense multitude to a few, whose possibility can either be supposed and postulated, or proved be experience.' \\

{\bf 1.2. Beables} \\

Enter John S Bell, [1], with the concept of 'beable' in Quantum Mechanics. \\

One of the aims Bell pursues with that concept is to discriminate among the multitude of 'observables' which are so
fundamental in the Copenhagen view of Quantum Mechanics, for instance. Indeed, a beable is supposed to be not merely
something which can be measured and registered as a real number, but only an observable which corresponds to a physical
reality, that is, it is independent of the act of observation. \\
Clearly, the very definition of beable cannot be seen as perfect, or in the above terms of Leibniz, as being conceived
through itself. Yet its importance and utility is no longer questioned among certain quantum physicists. \\
Furthermore, one should not omit noting that the concept of beable is theory dependent, that is, a given concept may be
a beable in a specific theory, and may cease to be so in another one. \\

{\bf 1.3. Modus Operandi} \\

It follows that {\it concepts} in Physics, as well as other sciences, can be seen as being of two rather distinct kind :
those which can clearly be defined in terms of certain other more basic concepts, and those which cannot, and therefore,
are by necessity taken for granted based on suitable intuitions, and possibly physical experiments. \\
Needless to say, such a division can depend on the way specific theories are set up. For instance, for Newton, gravitation
was a second type concept. \\

Upon the suggestion of Bell, and relevant not only to Quantum Mechanics, physical concepts are subjected to a further
division : those which are {\it beables}, and the rest. \\
And again, this division can depend on the particular buildup of one or another theory. \\ \\

{\bf 2. Configuration Spaces} \\

{\bf 2.1. A Widespread Disappointment} \\

As mentioned in [2, p. 5], one of the widespread arguments against pilot-wave theories comes from the following
disappointment : 'those sympathetic to the pilot-wave ontology no doubt expected that, for a system of N particles moving
in three spatial dimensions, the theoretical description would be of N wave-particle pairs – each pair consisting of a
point particle guided in some way by an associated wave propagating in 3-space. But Schroedinger's wave function for such
an N-particle system was emphatically not a set of N (interacting) waves, each propagating in 3-space. It was, rather, a
single wave propagating in the 3N-dimensional configuration space for the system.' \\
In this regard, it appears that Einstein himself was disappointed, that being a reason why he was not particularly
favourable towards the de Boglie-Bohm proposal of pilot-wave theory. \\

{\bf 2.2. Two Alternatives for the Composition of Systems \\
          \hspace*{0.85cm} of Quantum Particles} \\

Let us for convenience denote the expected {\it composite} situation by E-PHYS, namely that

\begin{itemize}

\item for a system of N quantum particles moving in three spatial dimensions, the theoretical description would be of N
wave-particle pairs – each pair consisting of a point particle guided in some way by an associated wave propagating in
3-space,

\end{itemize}

while what happens in fact as the actual {\it composition}, we denote by E-PROB, namely that
 
\begin{itemize}

\item Schroedinger's wave function for such an N-particle system is not a set of N, possibly interacting, waves, each
propagating in 3-space, but rather, a single wave propagating in the 3N-dimensional configuration space for the system.

\end{itemize}

In [2], the alternative E-PROB, which in fact happens, is rejected as not being physically meaningful. And then instead,
the rather involved theory in section 3, entitled 'complicated, ugly, and highly contrived' is suggested. \\

As far as we are concerned, we do not share that negative view of the author of [2], even if the respective method
suggested by him in the mentioned section leads to no less than a countable infinity of what one may call 'corrective
terms'. \\
Instead, we consider that the presence of those many corrective terms only indicates the price one must pay, if one
rejects the alternative which happens, namely, E-PROB, and instead, forces upon the situation the alternative E-PHYS. \\

Consequently, we argue here that the alternative E-PROB is supported by certain rather simple and natural physical
arguments, and thus it could be accepted, in which case the countably many corrective terms in section 3 of [2] need no
longer be necessary. \\

{\bf 2.3. Back to Basic Probability of Composite Systems} \\

Let us indeed consider the following simple situation in probability or statistics. We are given N random variables, be
they dependent or independent, namely \\

$~~~~~~ f_1 : E_1 \longrightarrow \mathbb{R}, \ldots, f_N : E_N \longrightarrow \mathbb{R} $ \\

where $E_1, \ldots, E_N$ are certain sample spaces. Then this system of N random variables is equivalent with the single
{\it composite} random variable \\

$~~~~~~ f : E \longrightarrow \mathbb{R}^N $ \\

where $E = E_1 \times \ldots \times E_N$, while $f ( x ) = ( f_1 ( x_1 ), \ldots, f_N ( x_N ) )$, for $x = ( x_1, \ldots,
x_N ) \in E$. \\

It follows, therefore, that the way of the composition E-PROB is accepted in probability and statistics as the one which
is relevant. \\

{\bf 2.4. Remembering the Born Rule} \\

The Born Rule has from the beginning of modern Quantum Mechanics played an important role, even if it may be seen in
more than one way. Indeed, let $S$ be a quantum system in a configuration space $E$, and let $X \subseteq E$ a Borelian
subset. If the state of the system $S$ is given at a moment in time by the wave function $\psi$, then the Born Rule states
that \\

$~~~~~~ Prob\, ( S \in X ) = \int_X |\, \psi ( x ) \,|\,^2 dx $ \\

Here, one can see $Prob\, ( S \in X )$ as the probability of the quantum system $S$ to be at the given time in the region
$X$ of the configuration space $E$, or merely, to be found there upon measurement. Also, according to certain views, the
Born Rule is not a beable, while according to other views it is. \\

An important aspect of the Born Rule, however, is that it is particularly useful in a better understanding of a number of
basic issues in Quantum Mechanics. From the start, for instance, it suggested, rightly or wrongly, that the wave functions
in the Schroedinger equation are {\it probability amplitudes}, rather than what Bell would call beables. Apart from that,
it can refer to many quantum processes, starting with some of the simplest ones, such as oscillators, scattering,
tunneling, and so on. \\

Be it, as it may, the Born Rule, in its own terms, is clearly about {\it probabilities}. Therefore, when it comes to its
application to {\it composite} quantum systems, it operates according to E-PROB, and {\it not} according to E-PHYS. \\ \\

{\bf 3. Back to Composition of Pilot-Waves of Quantum Systems} \\

The above may suggest that the widespread disappointment mentioned is subsection 2.1. above need not necessarily have a
basis. \\

Indeed, in pilot-wave theories, the composition of pilot-waves should rather be seen as given by E-PROB, than E-PHYS,
since those pilot-waves are supposed to come from the Schroedinger equation, thus in view of the Born Rule, their
composition may belong to the realm of probabilities, rather than to other realms. \\
And here, one should point to the fact that, whether one accepts the Born Rule, or not, whether one accepts it and sees it
as a beable, or not, so far there has not been any other view of it when accepted, except for the probabilistic one. \\

As for the Born Rule being problematic, so are a whole lot of other issues with quanta, and in fact, even with more
simple Physics ... \\
For instance, if a photon is indeed a wave, and the two slit experiment is done in void, then what is that which is
waving ? \\
Yes indeed, how can a wave, any wave for that matter, go through void ? \\
As for the simpler Physics of Newton, does gravitation propagate infinitely fast ? \\
And for the gravitational waves in General Relativity, they have not yet quite entered the realms of physical experience,
thus their possible status of beables is not so clear. \\

As for why and how interference happens in the two slit experiment done in void, claiming that it cannot be caused by
absolutely anything else but the superposition of two waves is a statement which is a mere declaration of faith, and at
that, of a faith based solely on non-quanta and classical type intuition ... \\

And then, let us suggest some minimal use of imagination in pilot-wave theories, one that seems not much more far fetched
than the interference of waves in void ... \\

Namely, the pilot-wave theories need not necessarily assume that both the particle and the wave are in the very same
configuration space. Indeed, one of the most basic enigmas which is utterly unanswered if we claim that they are both in
the very same configuration space is the following

\begin{itemize}

\item  How can a wave pilot a particle ?

\end{itemize}

So then, if the wave and the particle are in somewhat {\it different} configuration spaces, spaces which may still be
connected in certain yet mysterious ways, then N quantum particles have their joint configuration space built in the usual
manner. Namely, if for instance each of them is 3 dimensional, then their system is still the very same 3 dimensional
space. On the other hand, the corresponding N waves have their configuration space built as it happens with random
variables, that is, by Cartesian product, which gives a 3N dimensional space, as is the case with E-PROB.

\end{document}